\documentclass{optica-article}
\journal{opticajournal} 
\articletype{Research Article}
\usepackage{lineno}

\usepackage{caption}
\usepackage{subcaption}
\usepackage{array}
\usepackage{multirow}
\usepackage{color, soul}
\usepackage{float}
\usepackage[symbol]{footmisc}

\begin{document}

\title{Thermally induced refractive index trimming of visible-light silicon nitride waveguides using suspended heaters}

\author{Hong Chen,\authormark{1,4,$\dagger$} Tianyuan Xue,\authormark{1,2,6,$\dagger$} Zheng Yong,\authormark{2,5} Xianshu Luo,\authormark{3} Hongyao Chua,\authormark{3} Andrei Stalmashonak,\authormark{1} Guo-Qiang Lo,\authormark{3} Joyce K. S. Poon,\authormark{1,2} and Wesley D. Sacher\authormark{1,7}}

\address{\authormark{1}Max Planck Institute of Microstructure Physics, Weinberg 2, 06120 Halle, Germany\\
\authormark{2}Department of Electrical and Computer Engineering, University of Toronto, 10 King’s College Road, Toronto, Ontario M5S 3G4, Canada\\
\authormark{3}Advanced Micro Foundry Pte. Ltd., 11 Science Park Road, Singapore Science Park II, 117685, Singapore \\
\authormark{4}Now at: Xanadu, 777 Bay St., Toronto, Ontario M5G 2C8, Canada \\
\authormark{5}Now at: SCINTIL Photonics, Bâtiment BHT-3 69, rue Félix Esclangon, 38000 Grenoble, France\\
\authormark{6}tianyuan.xue@mpi-halle.mpg.de\\
\authormark{7}wesley.sacher@mpi-halle.mpg.de\\
\authormark{$\dagger$}These authors contributed equally.}

\begin{abstract*}
We demonstrate refractive index trimming of visible-light silicon nitride (SiN) waveguides using suspended heater structures. The thermal isolation of the suspended heaters enabled a semi-uniform temperature distribution with estimated temperatures of $\sim$350\textdegree C in the waveguides without reaching potentially damaging temperatures in the titanium nitride resistive heaters. The thermal isolation also enabled trimming temperatures to be reached with a moderate power dissipation of 30 to 40 mW. At a wavelength of 561 nm, modal effective index changes up to $-8.3 \times 10^{-3}$ were observed following thermal trimming, and the index changes were stable over an observation period of 97 days. The devices were fabricated as part of our visible-light integrated photonics platform on 200-mm diameter silicon wafers. The suspended heaters also functioned as efficient thermo-optic phase shifters with power dissipation for a $\pi$ phase shift of about $1.2-1.8$ mW. The trimming method was applied to set the bias points of thermo-optic Mach-Zehnder interferometer switches to reduce the bias power of five devices from $0.29-2.32$ mW to $0.1-0.16$ mW. Thermal trimming at a wavelength of 445 nm was also demonstrated. Through material analysis before and after thermal treatment, we hypothesize that index trimming of the silica (SiO$_2$) waveguide cladding may be a potential underlying mechanism. Additionally, via extrapolations of the measured trimming data, we estimate the thermal aging behavior of the SiN waveguides in the suspended heaters at lower (125 - 250\textdegree C) operating temperatures.
\end{abstract*}

\section{Introduction}
\label{sec:1:intro}
Photonic integrated circuits for the visible spectrum have the potential to enable transformative microsystems solutions for applications in neurotechnology \cite{Mohanty2020,Roszko2025}, biosensing \cite{kohler2021biophotonic}, quantum information \cite{niffenegger2020integrated,mehta2020integrated,Isichenko2023}, and beam forming \cite{poulton2017,mashayekh2021silicon,SharifAzadeh2023}. Achieving densely integrated photonic circuits (PICs) using wafer-scale production requires the use of silicon (Si) photonic foundry fabrication. Toward this, low-loss silicon nitride (SiN) \cite{Subramanian_2013,sacher2019visible} and aluminum oxide (Al$_{2}$O$_{3}$) \cite{west2019low,Al2O3_JSTQE_2019} waveguides for the visible spectrum have been demonstrated on 200- and 300-mm diameter Si wafers. Ongoing efforts seek to increase the complexity of visible-light photonic platforms; with SiN waveguides integrated with thermo-optic phase shifters \cite{Liang_Nature_Photonics_2021,Yong_OE_2022}, liquid crystal modulators \cite{notaros2018integrated,notaros2019liquid}, photodetectors \cite{Yanikgonul_Nature_Communications_2021,lin_NatureCommunications_2022}, and flip-chip bonded laser diodes \cite{MuOFC2025}. Recently, we have demonstrated a visible-light integrated photonics platform \cite{SacherOFC2023} with efficient bi-layer SiN edge couplers \cite{lin2021low}, Si photodetectors \cite{lin_NatureCommunications_2022}, electrothermal microelectromechanical (MEMS) cantilevers for beam steering \cite{SharifAzadeh2023}, and suspended thermo-optic phase shifters \cite{Yong_OE_2022}. One challenge with scaling these devices to large photonic circuits is the inherent sensitivity of visible-light integrated photonic devices to variations in waveguide dimensions and material refractive index --- a result of the relatively short wavelengths (relative to the near-infrared telecommunication wavelengths most often used in Si photonics, near 1310 and 1550 nm). Phase errors in waveguides due to fabrication variation affect the operating wavelengths of optical filters (e.g., arrayed waveguide gratings, ring resonators) and the bias points of interferometric switches/modulators such as Mach Zehnder interferometers (MZIs). These phase errors are often compensated using active phase tuning and power monitoring, with these functionalities also enabling locking of devices to varying laser wavelengths and stabilization in the presence of temperature fluctuations \cite{Dong:17}. A complementary approach to phase error compensation is post-fabrication refractive index trimming of waveguides, and numerous demonstrations exist of index trimming of Si \cite{schrauwen2008trimming,guo2016laser,milosevic2018ion,spector2016localized,jayatilleka2021post} and SiN (or silicon oxynitride) \cite{haeiwa2004wide,sparacin2005trimming,ueno2005high,xie2021post,de2020laser} waveguides for infrared wavelengths (often for devices targeting telecommunications applications). These trimming methods include electron radiation \cite{schrauwen2008trimming}, ultraviolet (UV) illumination \cite{haeiwa2004wide,ueno2005high,sparacin2005trimming,guo2016laser,de2020laser}, ion implantation \cite{milosevic2018ion,jayatilleka2021post}, and thermally annealed waveguide cores or cladding \cite{jayatilleka2021post,spector2016localized,xie2021post,Xie2023}. In principle, index trimming may be used in concert with active phase compensation methods (reducing the initial variability in device characteristics requiring active correction) or, in the case of PICs operating within temperature-controlled (e.g., cryogenic \cite{niffenegger2020integrated,mehta2020integrated}) environments, may be useful as a primary method of phase error compensation.  

Here, we propose and demonstrate refractive index trimming of visible-light SiN waveguides using monolithically integrated suspended heater structures fabricated in a visible-spectrum integrated photonics platform \cite{SacherOFC2023}. The thermal isolation of the suspended structure enables a semi-uniform temperature distribution with high temperatures in the SiN waveguides and SiO$_2$ cladding near 350$^{\circ}$C without reaching excessive temperatures in the titanium nitride (TiN) resistive heater --- for local index trimming of the SiN waveguides without damaging the TiN (Section \ref{sec:2:device}). The suspended heaters also function as efficient thermo-optic phase shifters, as we reported in Ref. \citenum{Yong_OE_2022}, with a $\pi$ rad. phase-shift power ($P_{\pi}$) of $\approx$ 1 mW at blue, green, and yellow wavelengths. The thermal isolation of the suspended heaters enables both low-power phase shifting and medium-power ($30-40$ mW) thermally induced index trimming (Section \ref{sec:3:results}), simplifying drive circuitry requirements when the same circuits are used for driving thermo-optic phase shifters and over-driving for \emph{in situ} thermal trimming of packaged PICs. Our simulations indicate that, without the suspended structure, the heater would require a power dissipation of about 950 mW to reach the same temperature used for trimming the SiN waveguides.

Additional measurements and analyses are presented to investigate the origin of the thermal trimming (Section \ref{sec:4:trimming_mech}) and, also, extrapolate the thermal aging behavior of the SiN waveguides at lower operating temperatures (Section \ref{sec:5:aging}). Based on Fourier transform infrared spectroscopy, we identify trimming of the SiO$_2$ waveguide cladding as a potential primary underlying mechanism. Independent of this hypothesis, we estimate modal effective index changes due to thermal aging of $\sim$ -1$\times 10^{-4}$ to -7.2 $\times10^{-3}$ for one year of operation and $\sim$ -2$\times 10^{-4}$ to -8.1 $\times10^{-3}$ for five years of operation at 125 - 250\textdegree C. These aging estimates are particularly relevant to applications of PICs requiring co-packaging or co-integration with electronic integrated circuits, wherein PICs may be exposed to temperatures $\gtrsim$ 100\textdegree C over multi-year operating lifetimes \cite{Kurata2021,Sahni:24}. For future investigation, we identify our thermal trimming method as a possible means of hardening devices against thermal and/or visible-light induced aging. This trimming method may also be useful for other wavelengths and materials, e.g., for trimming of SiN waveguides in Si photonics platforms for telecommunication wavelengths. We presented an initial report of these results in Ref. \citenum{SacherOFC2023} in addition to an initial report of this thermal trimming method at a wavelength of 1550 nm in Ref. \citenum{Xue2023}.

\section{Device design and fabrication}
\label{sec:2:device}

\begin{figure}
\centering
    {\includegraphics[width=0.9\textwidth]{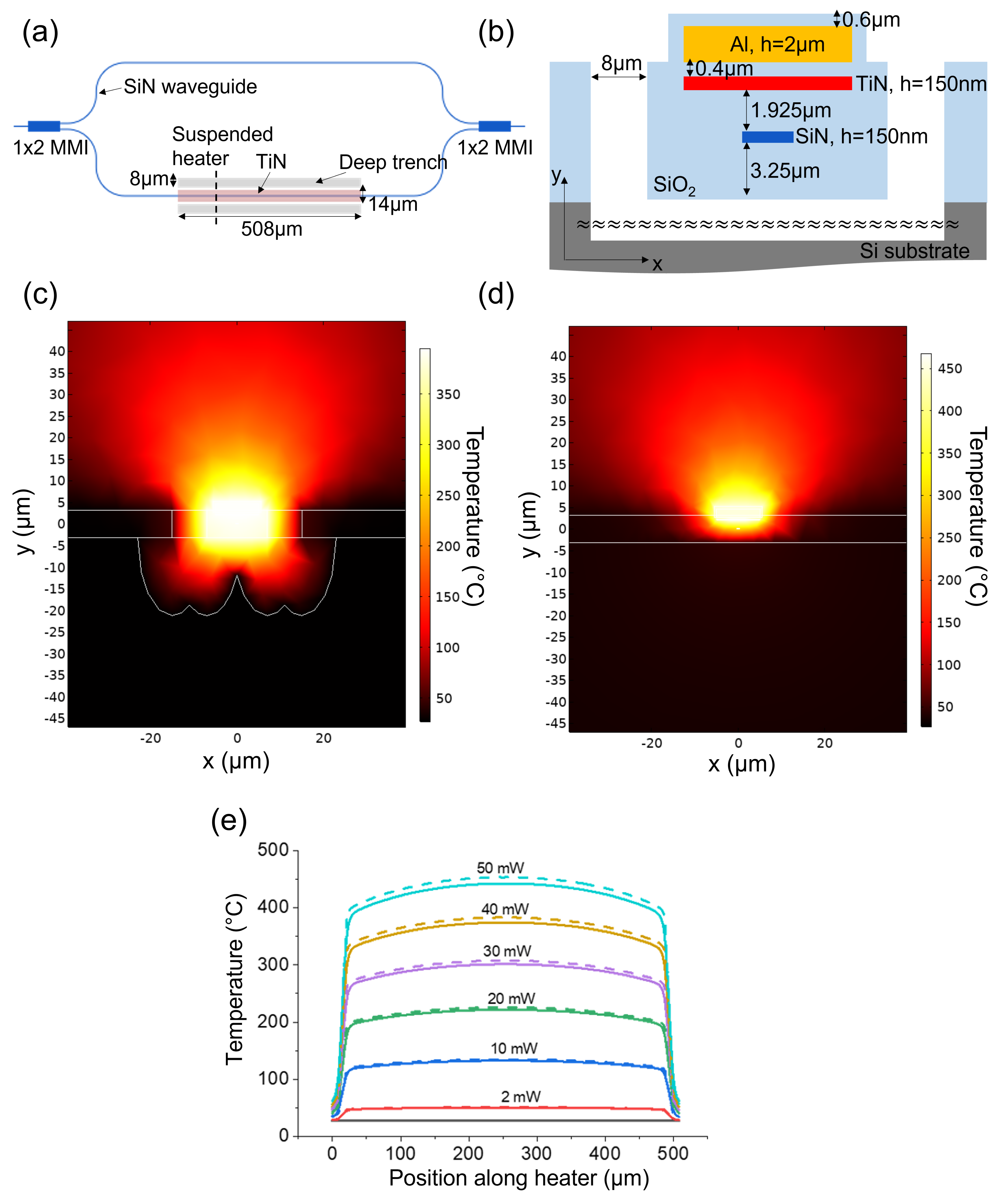} }
    \caption{Design and simulation of the suspended heater. (a) Schematic of the heater region in an AMZI test structure (not to scale); wires, vias, pads, and edge couplers are not shown. (b) Cross-sectional schematic (not to scale) corresponding to the dashed line in (a); layer thicknesses are denoted by $h$. (c) Simulated cross-sectional temperature profile within the suspended heater for 30 mW of applied power. The simulation volume exceeds the boundaries of the plot. (d) Simulated cross-sectional temperature profile of the heater without a suspended structure. To achieve a waveguide temperature similar to that in (c), 950 mW of power is required. (e) Longitudinal temperature profile through the center of the waveguide (solid line) and TiN heater (dashed line) along the suspended phase shifter at several heater powers.}\label{fig:fig1}
\end{figure}

The suspended heater design is shown in Fig. \ref{fig:fig1}(a). A thermally isolated suspended bridge was formed by two deep trenches and undercut etching of the Si substrate. A SiN waveguide and a TiN resistive heater were embedded in the suspended region, and the heating was localized to this region. TiN has a higher melting temperature (2950$^{\circ}$C) than aluminum (660\textdegree C) and copper (1083\textdegree C), and a high damage threshold temperature in excess of 700\textdegree C \cite{creemer2005mems} similar to tungsten (melting point of 3417\textdegree C) \cite{jayatilleka2021post}. The suspended structure dimensions were similar to our previously reported power-efficient thermo-optic phase shifters \cite{Yong_OE_2022}: the length of the suspended heater was 508 $\mu$m, and the widths of the deep trenches and the suspended silica (SiO$_2$) bridge were 8 and 14 $\mu$m, respectively. To simplify characterization, the structure was integrated into an asymmetric Mach-Zehnder interferometer (AMZI), with 1$\times$2 multimode interference (MMI) couplers at the input/output ports. A refractive index change in one of the AMZI arms results in a shift of the fringes in the transmission spectrum. The routing waveguide width was selected to be 600 nm for single-mode operation at a wavelength ($\lambda$) of $561$ nm. In addition, the MMI couplers were designed for minimum excess loss at $\lambda=561$ nm, and the length difference between the two AMZI arms was designed for a free spectral range (FSR) of 10 nm at this wavelength. 

The cross section of the suspended heater structure is depicted in Fig. \ref{fig:fig1}(b). The SiO$_2$ thickness between the TiN and SiN was designed to ensure negligible absorption loss \cite{Yong_OE_2022}. This structure was monolithically integrated into our visible-light integrated photonics platform, which includes a moderate-confinement SiN waveguide layer (150 nm thickness) for compact devices \cite{sacher2019visible}, a low-confinement SiN waveguide layer (75 nm thickness) for efficient bi-layer edge couplers \cite{lin2021low}, Si photodiodes \cite{lin_NatureCommunications_2022}, low-power thermo-optic phase shifters \cite{Yong_OE_2022}, and electrothermal MEMS devices \cite{SharifAzadeh2023}. The aluminum (Al) strip in the suspended structure, Fig. \ref{fig:fig1}(b), was not required for the device operation and was present since the device was originally designed for electrothermal-mechanical tests of suspended bridges --- later being repurposed for the thermal trimming experiments reported here.

Figure \ref{fig:fig1}(c) shows the temperature distribution in the suspended heater, simulated using COMSOL Multiphysics for 30 mW of applied electrical power. The simulation settings are detailed in Ref. \citenum{Yong_OE_2022}. The sheet resistance of TiN was assumed to be 10 $\Omega$/$\square$, and the Si substrate undercut profile was estimated from cross-section imaging. The simulated temperature distribution in the suspended region is semi-uniform, a result of the thermal isolation. To achieve similar temperatures in the SiN waveguide without the suspended geometry, 950 mW of applied electrical power would be required, as shown in Fig. \ref{fig:fig1}(d). Figure \ref{fig:fig1}(e) shows the temperature distribution along the optical propagation direction. The simulated temperature in the SiN waveguide reached about 350\textdegree C with 40 mW of applied power; the temperature in the TiN was only about 10\textdegree C higher. Outside the suspended heater, the temperature drops sharply --- presenting the possibility for trimming arrays of closely spaced devices. However, a detailed analysis is necessary to determine the density limits imposed by thermal crosstalk.

The devices were fabricated at Advanced Micro Foundry (AMF) on 200-mm diameter Si wafers. The fabrication process is described in Refs. \citenum{Yong_OE_2022} and \citenum{lin_NatureCommunications_2022}. Briefly, the fabrication process included: (1) Si mesa patterning and doping for photodiodes \cite{lin_NatureCommunications_2022}, (2) a series of SiO$_2$ and SiN plasma enhanced chemical vapor deposition (PECVD) steps, SiN waveguide patterning steps, and chemical mechanical polishing (CMP) to define the two SiN waveguide layers and the SiO$_2$ cladding, (3) TiN heater definition, (4) metallization steps to define two Al wiring layers and vias, (5) deep trench etching and Si substrate undercut etching to form facets for edge couplers and suspended structures. The SiN waveguide losses were reported in Ref. \citenum{Yong_OE_2022}.

\section{Measurement results}
\label{sec:3:results}

\subsection{Demonstration and characterization of thermal trimming}
\label{sec:3:1:trimming}

\begin{figure}[t!]
    \centering{\includegraphics[width=0.9\textwidth]{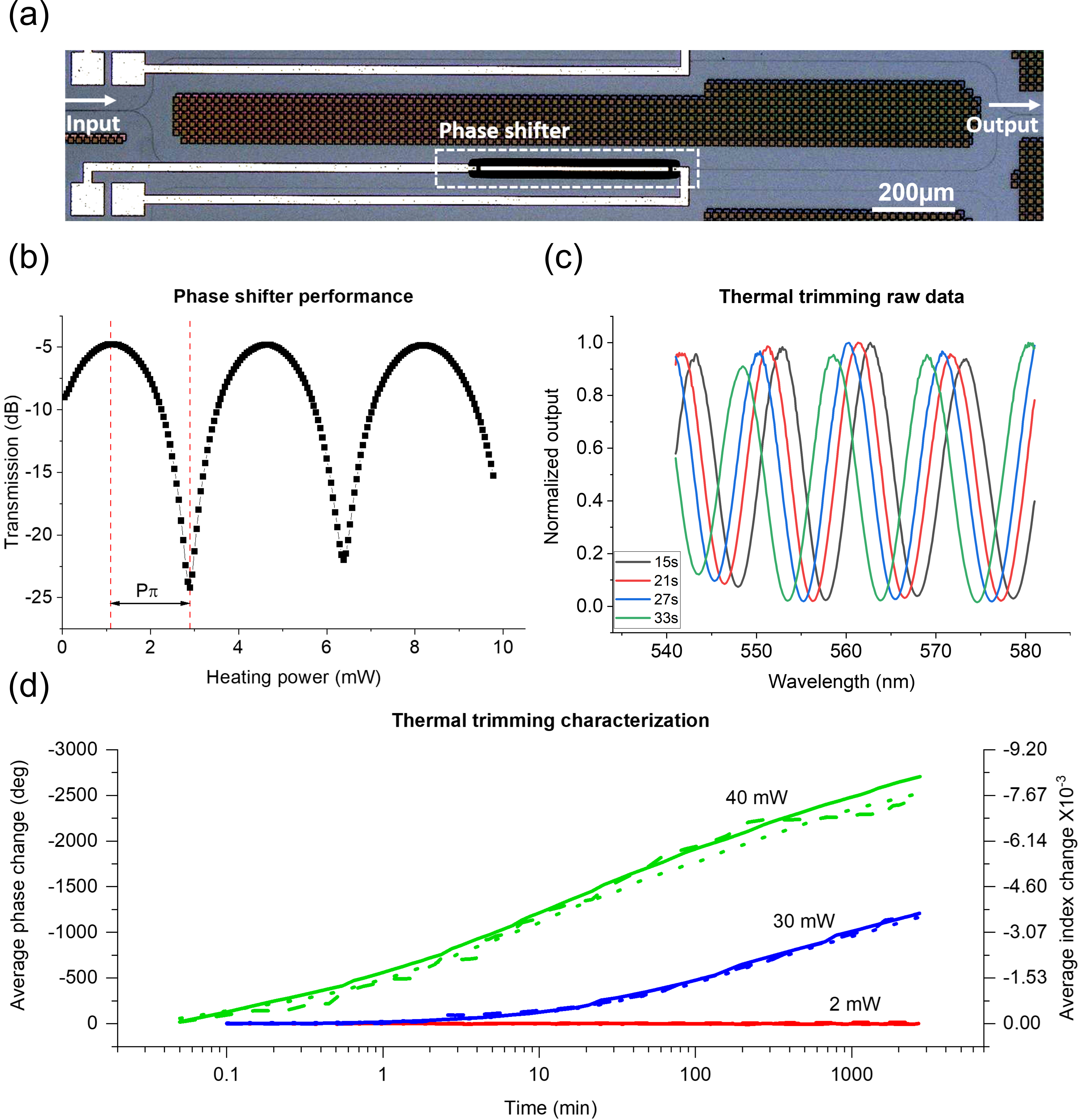} }
    \caption{Characterization of the thermal trimming. (a) Optical micrograph of a fabricated AMZI device with the suspended heater delineated by the white dashed box. (b) Measured transmission vs. heating power. (c) Measured transmission spectrum of the AMZI after 15, 21, 27, and 33 seconds of thermal trimming (40 mW heating power). (d) Thermal trimming experiment performed over an extended time period. Red, blue, and green curves correspond to heating powers of 2, 30, and 40 mW, respectively. For each heating power, three samples were randomly selected for trimming, and plotted with solid, dashed, and dotted traces.}\label{fig:fig2}
\end{figure}

A fabricated suspended heater integrated into an AMZI is shown in Fig. \ref{fig:fig2}(a). Laser light was edge-coupled to and from the AMZIs via cleaved single-mode optical fibers (Nufern S405-XP) mounted on piezoelectric-actuated micromanipulators (Thorlabs Nanomax MAX312D/M). The input optical polarization was selected to be transverse electric (TE) using an inline fiber polarization controller. The on-chip edge couplers had a nominal width of 5.2 $\mu$m at the facets, which was tapered to the routing waveguide width of 600 nm over a 300 $\mu$m length \cite{lin2021low}. The fiber-coupled output was directed to an optical power meter (Newport 818-SL/DB, 2936-R), and electrical power was applied to the heaters from a voltage supply (Keysight B2912A) via tungsten needle probes contacting the on-chip pads. 

To characterize the thermal trimming, modal effective index changes in the suspended heater structure were extracted through measurements of the AMZI transmission spectrum. Here, we used a supercontinuum laser (SuperK FIANIUM FIU-15) coupled to a wavelength-tunable optical filter (Photon Etc. LLTF Contrast VIS) as the input light source to the measurement setup. Transmission spectra were collected by sweeping the filter wavelength while measuring the output optical power of the AMZI. The full-width at half-maximum (FWHM) linewidth of the filtered laser output was about $1-2$ nm, limiting the resolution of the measured transmission spectra. The resolution was sufficient for tracking shifts of the fringes of the AMZI spectra, but the troughs of the transmission spectra were only partially resolvable. To determine the phase shifter power consumption and AMZI insertion loss and extinction ratio, we measured the change in optical transmission as a function of applied electrical power (at a fixed wavelength) using a single-wavelength laser at $\lambda = 561$ nm (Coherent Sapphire FP). The relatively narrow linewidth of this laser allowed us to resolve the extinction ratio of the AMZI and $P_{\pi}$ of the suspended heater. 

A typical AMZI transmission versus applied electrical power trace is presented in Fig. \ref{fig:fig2}(b), and the measured insertion loss and extinction ratio were 4.8 and 19.4 dB, respectively. The AMZI insertion loss was determined by normalizing the fiber-coupled output power to the output of a neighboring reference 600-nm-wide waveguide sharing the same edge coupler design. Thus, the AMZI insertion loss consists of the losses from the MMI couplers, suspended heater phase shifter, and waveguide propagation. From cutback measurements of waveguide loss, we estimate the loss of the phase shifter to be $< 0.3$ dB. The applied power to the heater for a $\pi$ phase shift, $P_{\pi}$, was 1.77 mW. 

Figure \ref{fig:fig2}(c) shows the spectra of the AMZI at several times during the thermal trimming process with 40 mW of applied power to the TiN heater, corresponding to an estimated temperature of 350\textdegree C in the suspended region (as simulated in Section \ref{sec:2:device}). The input wavelength was tuned by the optical filter coupled to the supercontinuum laser. At each of the times in Fig. \ref{fig:fig2}(c), the trimming process was paused, i.e., the heater power was disconnected and the device was allowed to cool for 1-2 minutes, and then an optical spectrum was collected. From 15 s to 33 s of total trimming time, the interference fringes shifted with the increasing trimming duration, indicating that the modal effective refractive index changed in the suspended region. The blue shift of the spectrum indicates a negative index change, and negative index changes were consistently observed throughout this work. To extract the phase and effective index changes from optical transmission measurements, we note that the total transmitted output power of the AMZI follows:
\begin{equation}
I_{out} = \frac{I_{in}}{2}[1+\cos(\Delta\phi)]
\end{equation}
where $I_{in}$ is the total optical power in the two arms of the AMZI and $\Delta\phi$ is the phase difference between the two arms of the AMZI. Furthermore, $\Delta\phi=\frac{2\pi}{\lambda}n_{eff} \Delta L$, where $\lambda$ is the wavelength, $\Delta L$ indicates the length difference between the two arms, and $n_{eff}$ denotes the modal effective index. After trimming, the phase difference between the two arms will be changed and can written as:
\begin{equation}
\Delta\phi = \frac{2\pi}{\lambda} n_{eff} \Delta L + \theta
\end{equation}
where $n_{eff}$ is the effective index before trimming and $\theta$ is the additional optical phase change in the trimmed arm (referred to hereafter as the \emph{phase change}), which is related to the effective index change by $\theta=\frac{2\pi}{\lambda} \Delta n_{eff} L_{trim}$. $L_{trim}$ is the length of trimming section, and in the following analysis, we assume $L_{trim}=$ 508 $\mu$m, the length of the suspended heater region. By sampling the output power at a fixed wavelength during trimming, the phase change and corresponding effective index change can be calculated.

Figure \ref{fig:fig2}(d) shows trimming over an extended time period by automating the test and data acquisition. The trimming was performed with the heater power dissipation ranging from 2 to 40 mW over $\approx$2700 minutes. Spectral shifts were recorded with uneven time steps of a few seconds in the initial stage of trimming, to 3 hours in the last stage of trimming. The measurement time step was chosen to ensure the phase change between successive measurements was less than 2$\pi$. The trimming process was paused and the device was allowed to cool before each measurement. For each applied electrical power, three devices (from three different chips) were selected from the wafer for trimming measurements. The maximum effective index change was observed when 40 mW of power was applied for 2738 minutes, and the absolute modal effective index change averaged over the length of the phase shifter ($|\Delta n_{eff,avg}|$) was 8.3$\times$10$^{-3}$ (averaged over the three devices). With shorter trimming times of 1 and 10 min, $|\Delta n_{eff,avg}|$ was 1.5$\times$10$^{-3}$ and 3.6$\times$10$^{-3}$, respectively. When 30 mW of power was applied, the rate of the index change was lower, and after 2713 minutes, $|\Delta n_{eff,avg}| =  3.7 \times$10$^{-3}$ (averaged over three devices). For the six samples measured at 30-40 mW of heater power, the thermal trimming was repeatable with variations in $|\Delta n_{eff,avg}|$ of 3.1\% (30 mW) and 4.6\% (40 mW) over the $\approx$2700 minutes of trimming; the magnitude of variations in $|\Delta n_{eff,avg}|$ between samples at fixed heater power did not appear to change significantly with trimming time.

For 2 mW of applied electrical power (similar to $P_{\pi}$), the measured phase shift observed was within $20^\circ$ of its initial value, corresponding to an effective index change of 5$\times$10$^{-5}$. This low phase shift may be due to ambient temperature drift and measurement error. When the temperature of the stage supporting the photonic chips was controlled, the phase shift fluctuated near zero with a maximum amplitude of 12\textdegree. In addition to the applied powers in Fig. \ref{fig:fig2}(d), other powers were tested and the index change was found to be negligible when the applied power was $\leq 10$ mW, corresponding to a simulated temperature of $\leq 132$ $^{\circ}$C in the suspended region. This confirms that, with low applied powers, the suspended heaters may also operate as thermo-optic phase shifters without trimming of the waveguides.

\subsection{Application: Bias power reduction of thermo-optic MZI switches}
\label{sec:3:2:trim2bias}
As an example application, we applied the thermal trimming method to adjust the bias points of 1$\times$2 thermo-optic Mach-Zehnder interferometer (MZI) switches. Fabrication variations lead to phase errors between the MZI arms, and thus, device-to-device variations in the ratio of the outputs. A bias power can be applied to the phase shifter to compensate for the phase error, at the expense of increased power consumption. Here, we use thermal trimming to compensate for the MZI phase error, and target maximum/minimum transmission at the output ports with no power applied to the phase shifter --- reducing the bias power required for the MZI switch.

\begin{figure}[hb!]
\centering
    \centering{{\includegraphics[width=0.9\textwidth]{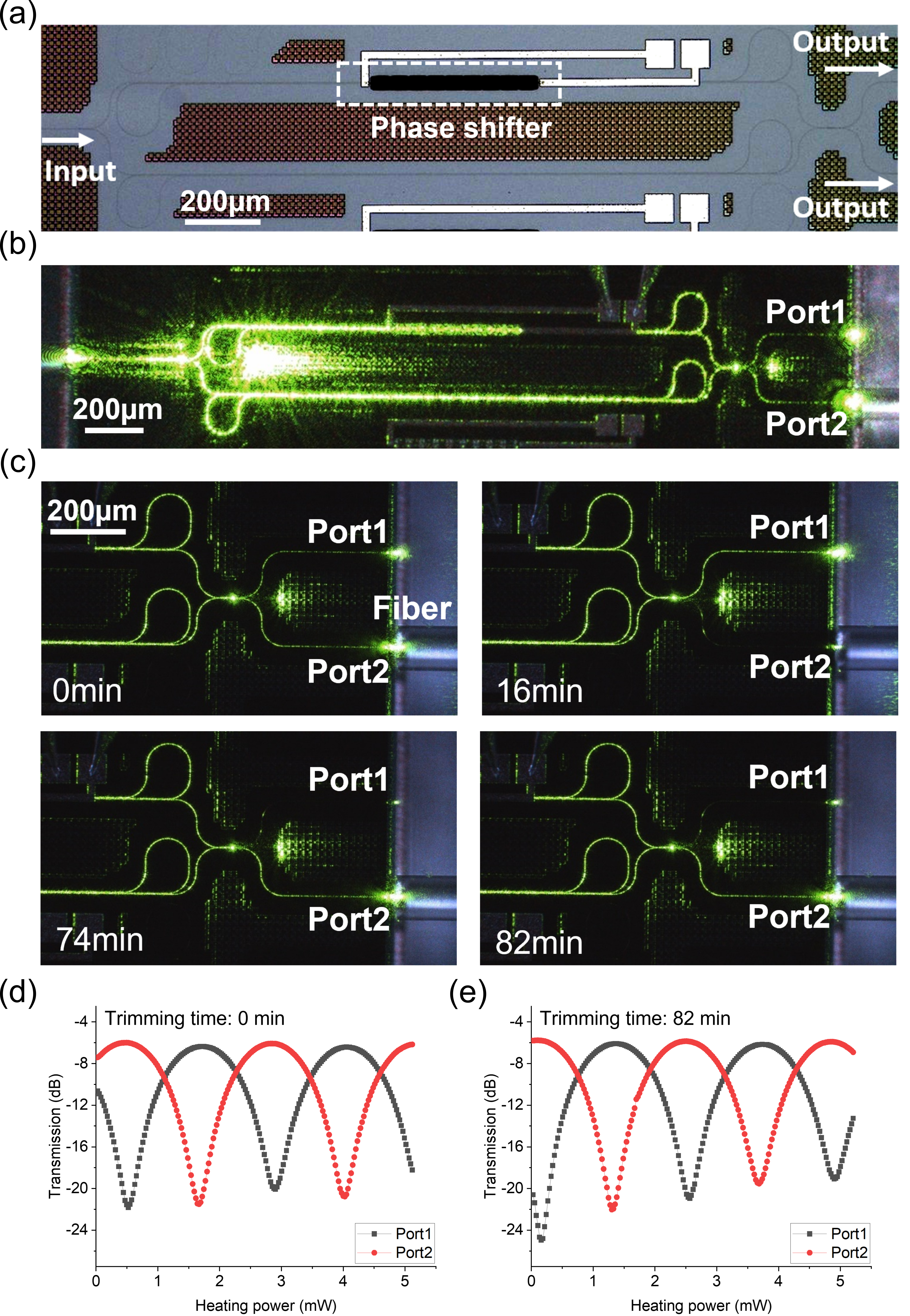} }}
    \caption{Thermal trimming of a 1$\times$2 MZI switch with a multi-pass, suspended, thermo-optic phase shifter. (a) Optical micrograph of the fabricated device. (b) Optical micrograph of the device with input laser light at $\lambda = 561$ nm. (c) Magnified views of the MZI outputs at various times throughout the trimming process. (d),(e) Transmission versus heating power (d) before trimming and (e) after 82 minutes of trimming. The heater power was disconnected and the device was allowed to cool for 1-2 minutes before collecting the data in (c) - (e).}\label{fig:fig3}
\end{figure}

The MZI switch had a multi-pass, suspended, thermo-optic phase shifter in one arm; we previously reported the design in Ref. \citenum{Yong_OE_2022}. The phase shifter had 3 waveguides passes under the TiN heater in the suspended region. An optical micrograph of the device is shown in Fig. \ref{fig:fig3}(a). The MZI had a 1$\times$2 MMI coupler at the input side, a 2$\times$2 MMI coupler at the output side, and the arms were of nominally equal path length. The measured switching power, $P_{\pi}$, was 1.2 mW at $\lambda=561$ nm. The single-wavelength laser (Coherent Sapphire FP) was used for measurements in this subsection. We reported detailed characterization of the device in Ref. \citenum{Yong_OE_2022}. Figure \ref{fig:fig3}(b) shows the device transmitting $\lambda=561$ nm light before trimming, and Fig. \ref{fig:fig3}(c) (top-left) shows a magnified view of the outputs (Port1 and Port2). Without any applied heater bias, the optical powers at the two output ports were similar, and an applied power of 0.52 mW was required to minimize the output power in Port1, as shown in the transmission versus heating power traces in Fig. \ref{fig:fig3}(d). The reported transmission corresponds to the insertion loss of the MZI excluding the edge couplers.

Figure \ref{fig:fig3}(c) also shows optical micrographs of the MZI outputs at various times during thermal trimming with 36 mW of applied power. After 16 minutes, the output was minimized in Port2 and maximized in Port1. The output was maximized in Port2 and minimized in Port1 after 74 minutes. The transmission versus heating power after 82 minutes of thermal trimming is plotted in Fig. \ref{fig:fig3}(e); the bias power to minimize the Port1 output power was reduced to 0.16 mW. Similar to Section \ref{sec:3:1:trimming}, the heating power was disconnected and the device was allowed to cool for 1 - 2 minutes prior to collecting the micrographs and transmission data in Fig. \ref{fig:fig3}. The trimming process was monitored by pausing the trimming process and collecting transmission vs. heating power traces at various time intervals throughout the process; the intervals decreasing in length as the Port1 transmission approached its minimum.

\begin{figure}[h]
    \centering {\includegraphics[width=1\textwidth]{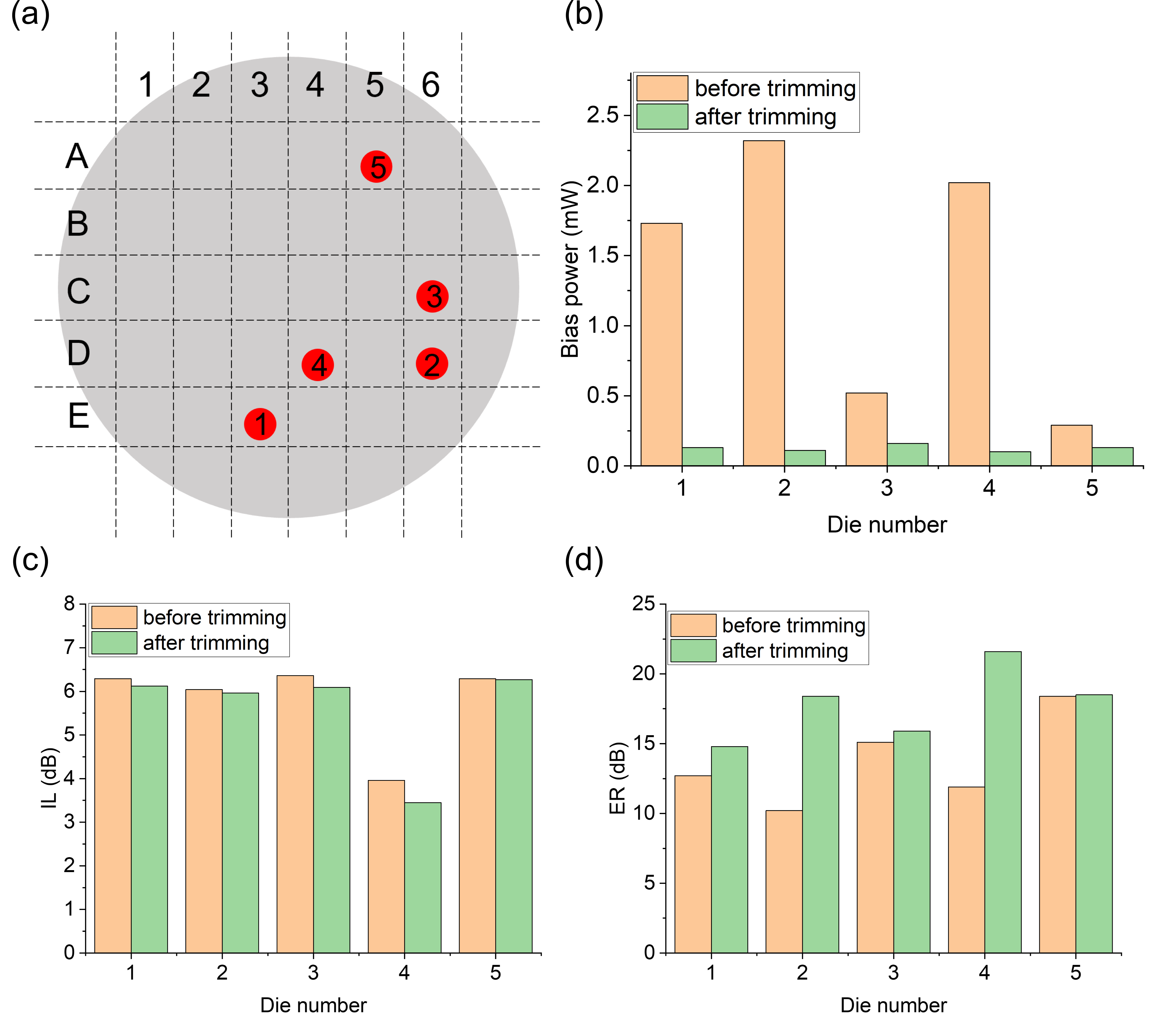} }
    \caption{Trimming results for thermo-optic MZI switches across a wafer ($\lambda = 561$ nm). (a) Locations of the measured dies on the wafer. (b) Bias heater power to minimize the output power in Port1, (c) MZI insertion losses (IL), and (d) MZI extinction ratios (ER) before and after trimming.}\label{fig:fig4}
\end{figure}

In total, the measurements were performed for 5 devices on different dies from the wafer as shown in Fig. \ref{fig:fig4}(a); the device in Fig. \ref{fig:fig3} was sample number 3. For each MZI, the bias power, insertion loss, and extinction ratio were recorded before and after trimming. The bias power is defined to be the minimal heater power with which the Port1 transmission is minimized. As summarized in Fig. \ref{fig:fig4}(b), due to fabrication variation in the SiN waveguide dimensions, the initial bias power ranged from 0.29 to 2.32 mW. Following thermal trimming, the bias power was reduced to 0.10 to 0.16 mW.

The measured insertion losses of the MZIs on dies 1 to 5 before(after) trimming were 6.3(6.1), 6.0(6.0), 6.4(6.1), 4.0(3.5), 6.3(6.3) dB as shown in Fig. \ref{fig:fig4}(c). These values are consistent with our previous report \cite{Yong_OE_2022}. After trimming, the insertion losses were slightly reduced on average. This may be related to a reduction in defects or a modified morphology of the SiO$_2$ cladding and/or SiN, and possible underlying mechanisms for the thermal trimming effects are discussed in Section \ref{sec:4:trimming_mech}. Figure \ref{fig:fig4}(d) also shows that the extinction ratios improved with the trimming. We hypothesize that the MZI arm with the phase shifter had a slightly higher loss than the other arm, and the waveguide loss reduction from trimming partially compensated for this imbalance. A similar trimming test of a 1$\times$2 MZI switch with blue ($\lambda =$ 445 nm) light was also performed (see Appendix). Possible improvements to this trimming method --- for increased speed and accuracy --- include pulsed trimming, on-chip power monitoring, and optimized cooling times between trimming and measurement steps.

\subsection{Repeatability and stability of trimming}
\label{sec:3:3:stability}

Our quantification of the index trimming was based on the AMZI measurements in Section \ref{sec:3:1:trimming}. Since the trimming of each device was investigated over extended time periods up to about 2700 minutes (with many spectra collected throughout this time), multiple factors contributed to the measurement error: (1) temperature fluctuations in the environment; (2) polarization drifts in the fiber; and (3) mechanical drifts of fiber-to-chip alignment that added fluctuating Fabry-Perot fringes. Moreover, due to the $\approx$ 1-2 nm linewidth of the wavelength-tunable optical filter used to measure optical spectra, the MZI spectrum (FSR $\approx$ 9 nm) was significantly averaged. By measuring AMZIs without any applied electrical power, we observed a drift of the AMZI transmission spectra of about $\pm 0.5$ nm, corresponding to a measurement error of about $\pm 20^\circ$.

\begin{table}[!t]
\caption{List of thermo-optic phase shifters characterized immediately after trimming and after an extended period of time in storage, sample labeling is defined in Fig. \ref{fig:fig4}(a)} 
    \centering
    \begin{tabular}{c | c | c | c | c | c}
    \hline
         \begin{tabular}{@{}c} Wafer \\ number \\ \end{tabular} &  
        \begin{tabular}{@{}c} Sample \\ number \\ \end{tabular} & 
        \begin{tabular}{@{}c} Trimming \\ power \\ (mW) \end{tabular} & 
        \begin{tabular}{@{}c} Phase \\ change (deg) \\ \end{tabular} &
        \begin{tabular}{@{}c} Days between \\ measurements \\ \end{tabular}&
        \begin{tabular}{@{}c} Relaxation \\ of phase \\ change (deg) \end{tabular}\\[0.5ex] 
    \hline
    \multirow{2}{*}{Wafer 1} & B1 & 2 & -3 & 97 & 12 \\
    \cline{2-6}
    & A3$^a$ & 2 & -18 & 100 & 4 \\
    \cline{2-6}
    & C4$^b$ & 2 & 12 & 103 & -4 \\
    \cline{2-6}
    & E2$^a$ & 10 & 1 & 106 & 25 \\
    \cline{2-6}
    & C2 & 30 & -1118 & 97 & 32 \\
    \cline{2-6}
    & D1 & 30 & -1209 & 103 & 10 \\
    \cline{2-6}
    & D2$^a$ & 30 & -1207 & 107 & 21 \\
    \cline{2-6}
    & E3$^a$ & 30 & -1167 & 99 & 10 \\
    \cline{2-6}
    & B4$^{a,b}$ & 40 & -2545 & 99 & 3 \\
    \cline{2-6}
    & E5$^{a,b}$ & 40 & -2420 & 113 & 7 \\
    \cline{2-6}
    & D5$^a$ & 40 & -2705 & 100 & 8 \\
    \cline{1-6}
    \multirow{3}{*}{Wafer 2} & A6 & 10 & 20 & 106 & 32\\
    \cline{2-6}
    & C2 & 20 & -93 & 118 & 52\\
    \cline{2-6}
    & D5 & 40 & -1809 & 106 & 16\\
    \cline{2-6}
    & E3 & 40 & -2109 & 115 & 41\\
    \hline
    \end{tabular}
    $^a$storage in rough vacuum\\
    $^b$measurements on temperature-controlled stage 
    \label{table:trim}
\end{table}

To test the repeatability and long-term stability of the trimming, multiple AMZIs were measured immediately after trimming (trimming time $\approx$ 2700 minutes) and after about three months in storage (using the method in Section \ref{sec:3:1:trimming}). The results are summarized in Table 1. Dies from two wafers were measured: Wafer 1 corresponds to the cross-section in Fig. \ref{fig:fig1}(b) and the measurement results in Section \ref{sec:3:results}, and Wafer 2 is nominally identical to Wafer 1 with the exception that the bottom cladding of the SiN waveguides was 250 nm thinner. For trimming powers below 10 mW, immediately after trimming, phase changes varying from -20\textdegree\ to 18\textdegree\ were observed, and those values were within the measurement error range. While for trimming powers of 30-40 mW, large phase changes were measured immediately after trimming; for Wafer 1, -1209\textdegree\ to -1118\textdegree\ (30 mW) and -2705\textdegree\ to -2420\textdegree\ (40 mW). The difference in phase changes was larger between Wafers 1 and 2 rather than within each wafer. This may be due to fabrication variation between the wafers, e.g., differing SiN thickness and waveguide widths (affecting the optical mode confinement), and variations in deep trench and undercut etching (affecting the temperatures in the suspended heaters at a given trimming power).

Following about three months in storage at roughly room temperature, the samples in Table \ref{table:trim} were remeasured. The relaxation of the phase change is the difference between these remeasured phase changes and the phase changes measured immediately after trimming. Generally, the relaxation of the phase change was small (-41\textdegree\ at worst for 30-40 mW of trimming power) and comparable to the measurement error. However, it is noteworthy that the difference in the phase change is also generally opposite to the phase change due to trimming, indicating that a slight relaxation of the trimming occurred, and this cannot be attributed entirely to measurement error.

Three of the dies in Table 1 (Wafer 1 - C4, B4, E5) were measured using a temperature-controlled stage, reducing the impact of ambient thermal fluctuations on the measurements. The magnitude of the measured phase relaxation is smaller in these cases, a result of the reduced measurement error. In addition, seven out of the 11 Wafer 1 dies were stored in a rough vacuum, however, large differences in the phase relaxation were not observed between the samples stored in vacuum versus atmospheric pressure.

\section{Investigation of trimming mechanism}
\label{sec:4:trimming_mech}

\begin{figure}[!ht]
    \centering
    \includegraphics[width=0.92\textwidth]{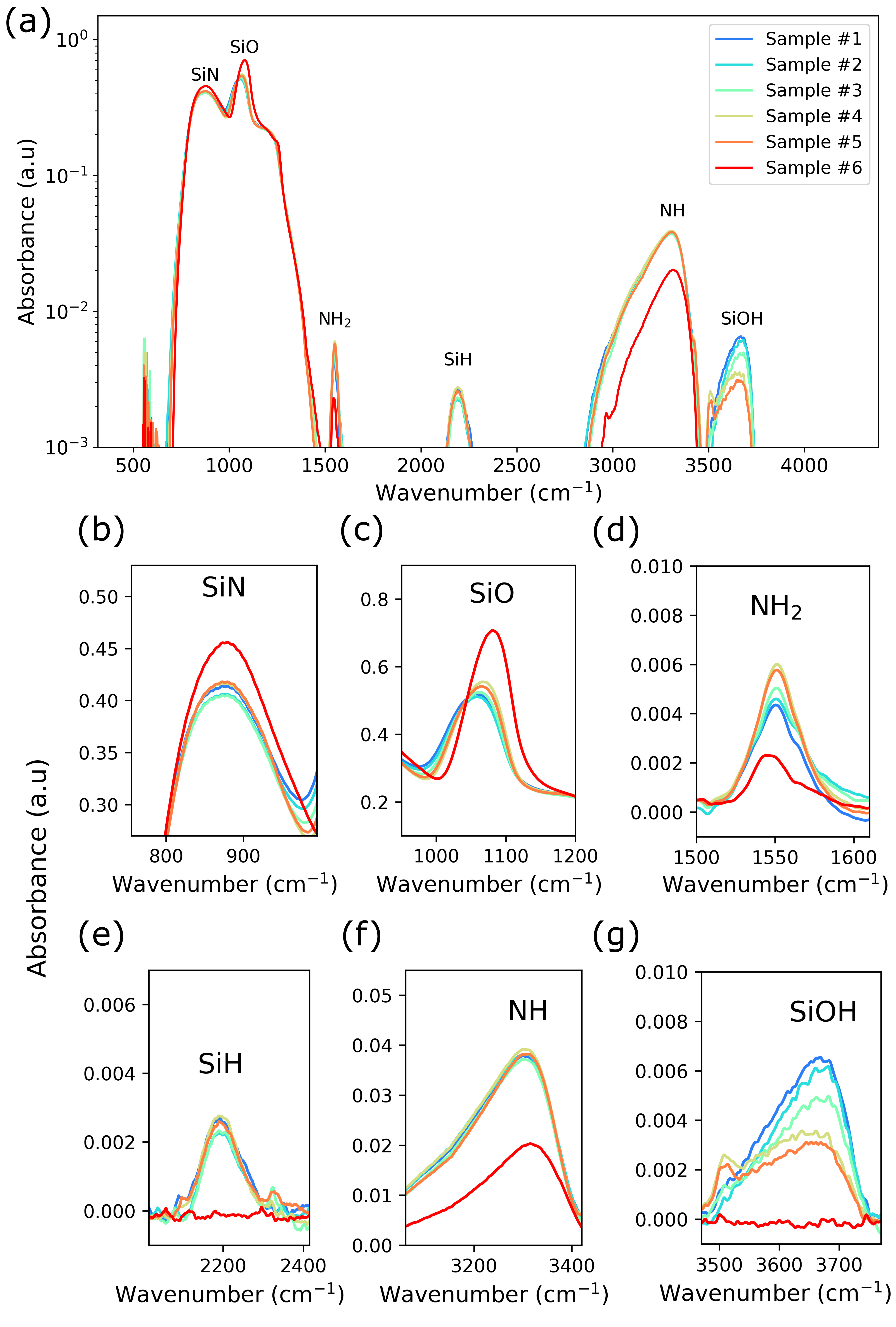}
            \caption{Transmission FTIR spectroscopy of Samples \#1-6 processed under annealing parameters indicated in Table \ref{table:bake}. (a) Measured FTIR spectra. (b)-(g) Zoomed-in views of peaks in (a) associated with (b) SiN bonds, (c) SiO bonds, (d) NH$_2$ bonds, (e) SiH bonds, (f) NH bonds, and (g) SiOH bonds.}
            \label{fig:FTIR}
\end{figure}

To investigate the origins of the observed thermal trimming of the SiN waveguides, material analyses of the SiN waveguide cores and SiO$_2$ cladding were performed. First, transmission electron microscopy (TEM) and energy dispersive X-ray (EDX) analysis indicated that the waveguide core was composed of silicon and nitrogen, with negligible oxygen content --- confirming that the waveguide core material was SiN rather than silicon oxynitride (SiON). Similarly, the analysis showed that the cladding was composed of silicon and oxygen, with negligible nitrogen content, which confirmed that the cladding was SiO$_2$ and not SiON. Next, a "blank wafer" consisting of 400 nm of SiN (without patterning) atop 400 nm of SiO$_2$ on a Si substrate was prepared (the thin films deposited via the same processes used in our visible-light integrated photonics platform). Ellipsometry was performed on samples from the blank wafer, and the measured SiN refractive index was about 1.9 at $\lambda = 561$ nm --- lower than the reported index of stoichiometric Si$_3$N$_4$, indicating a nitrogen rich SiN film \cite{bucio2016material}. Also, transmission Fourier transform infrared (FTIR) spectroscopy was performed on 6 cleaved chips from the blank wafer with and without subsequent thermal treatment (annealing) in a tube furnace (at varying target temperatures and durations, see Table \ref{table:bake}). The measured FTIR spectra for each of the samples are shown in Fig. \ref{fig:FTIR}; a reference sample consisting of only the Si substrate was used to remove the signal from the substrate.

\begin{table}[t!]
\caption{Annealing parameters for samples analyzed by FTIR spectroscopy. Tube furnace ramp rate for all annealed samples was 10\textdegree C/min} 
    \centering
    \begin{tabular}{c | c | c }
    \hline
         \begin{tabular}{@{}c} Sample number \end{tabular} &  
        \begin{tabular}{@{}c} Target temperature \end{tabular} & 
        \begin{tabular}{@{}c} Duration held at target temperature \end{tabular}\\[0.5ex] 
    \hline
    1 & Not annealed & Not applicable \\
    \hline
    2 & Not annealed & Not applicable \\
    \hline
    3 & 350\textdegree C & 4 hours \\
    \hline 
    4 & 450\textdegree C  & 2 hours \\
    \hline
    5 & 430\textdegree C  & 3 hours \\
    \hline
    6 & 1000\textdegree C  & 4 hours \\
    \hline
    \end{tabular}
    \label{table:bake}
\end{table}

Samples 3-5 were exposed to annealing temperatures comparable to the estimated 300 - 400\textdegree C trimming temperatures in this work. In Fig. \ref{fig:FTIR}, the FTIR spectra of these samples show that the peaks associated with the SiN (871 cm$^{-1}$ \cite{Ay2004}), SiH (2198 cm$^{-1}$ \cite{Ay2004}) and NH (3326 cm$^{-1}$ \cite{Ay2004}) had small to negligible differences after thermal treatment and no discernible trend with the annealing temperature and duration. The evolution of the peaks in the FTIR spectroscopy data displaying a consistent trend with increasing annealing temperature and duration (from Sample 1-5) are as follows. SiO ($\approx$ 1060 cm$^{-1}$ \cite{Ay2004}) underwent a progressive blue-shift and increase in intensity, Fig. \ref{fig:FTIR}(c). NH$_2$ (1543 cm$^{-1}$ \cite{Stann2002}) progressively increased in intensity, Fig. \ref{fig:FTIR}(d), and SiOH ($\approx$ 3650 cm$^{-1}$ \cite{Ay2004}) progressively decreased in intensity, Fig. \ref{fig:FTIR}(g).

Sample 6 was exposed to a considerably higher annealing temperature (1000\textdegree C) than the estimated trimming temperatures used in this work. Compared to the Samples 1-5, the FTIR spectrum of Sample 6 exhibited a continuation of the general trends (with increasing annealing temperature) in the peaks associated with SiO, NH$_2$, and SiOH. In addition, large reductions in the intensities of the peaks associated with SiH and NH occurred in Sample 6 (with no clear reduction in these peaks observed at the lower temperatures of Samples 3-5). Qualitatively, we also observed that the SiO peak changed much more with exposure to 1000\textdegree C (Sample 6) compared to 350-450\textdegree C (Samples 3-5), while both temperatures were able to significantly reduce the SiOH peak intensity. 

From the outset, the underlying mechanism of the thermal trimming in this work may be related to refractive index changes in the SiN waveguide cores, SiO$_2$ cladding, or both. 
In the remainder of this discussion, we consider these possibilities together with the above material analysis results and the observation from Section \ref{sec:2:device} that the thermal trimming generally resulted in negative effective index changes in the SiN waveguides. 
First, considering changes in the SiN, breaking of SiH and NH defect bonds \cite{de2020laser} are a known mechanism for refractive index changes. 
Prior studies of SiN thermal annealing in Refs. \citenum{Cai1996,Yoo2007,Wright2008,Parkh2023,Lavareda2023,Blasco2024,Kobayashi2024} reported an increase of refractive index of SiN with annealing, and the effects were most pronounced for Si-rich and stoichiometric SiN (unlike the nitrogen-rich SiN in this work). 
Reference \citenum{Mei2006} reported a reduction in refractive index in nitrogen-rich and slightly Si-rich SiN films, but the annealing temperature (690\textdegree C) was much higher than the trimming temperatures in this work. 
Overall, these prior studies do not clearly point to SiN annealing as the underlying mechanism for the negative effective index changes with thermal trimming (at $\approx$ 300 - 400\textdegree C) observed here.

Next, considering changes in the SiO$_2$ cladding during trimming, prior annealing studies of SiO$_2$ thin films deposited through plasma-assisted processes report decreasing refractive index with annealing temperatures < 600\textdegree C \cite{Adams1981,Fitch1990,Haque1997,Ghaderi2016} and increasing refractive index for higher temperatures \cite{Pan1985,Fitch1990,Parkh2023}. 
This transition has been explained as two separate regimes \cite{Fitch1990} where: (1) at lower temperatures (< 600\textdegree C), a release of hydrogen from SiOH bonds occurs which forms nanopores in the material, and (2) at higher temperatures (> 600\textdegree C), structural changes in SiO$_2$ occur and the material densifies \cite{Bigl2015,Fu2017,Uedono2024}.
The reduced density of the SiO$_2$ cladding through the formation of nanopores leads to a decrease in the refractive index via the Clausius-Mossotti relation.
From optical mode simulations of the waveguide, we find that a change in the SiO$_2$ cladding refractive index of -0.019 corresponds to the -8.3$\times$10$^{-3}$ change in effective modal index observed in this work.

Considering our material analysis results together with the above-mentioned prior studies, we hypothesize that refractive index changes in the SiO$_2$ cladding may potentially be a primary mechanism underlying our thermal trimming demonstrations.
The negative waveguide effective index changes we observed with thermal trimming in addition to the reduction in the measured FTIR spectral peak associated with SiOH, Fig. \ref{fig:FTIR}(g), are consistent with the expected negative index change of SiO$_2$ annealed at < 600\textdegree C \cite{Adams1981,Fitch1990,Haque1997,Ghaderi2016} and the reported mechanism of thermal decomposition of SiOH bonds \cite{Fitch1990}. 
We further hypothesize that changes in the SiH and NH bonds and structural changes in the SiO$_2$ occur at temperatures beyond the trimming temperatures used here. 
This is consistent with our observation that large changes in NH, SiH, and SiO FTIR spectral peaks occurred only in the high-temperature-annealed Sample 6. 
The behavior of the NH$_2$ FTIR spectral peak (increasing in intensity for Samples 3-5 and decreasing for Sample 6) may be due to, first, diffusion of hydrogen from the SiO$_2$ cladding into the SiN following decomposition of SiOH bonds at lower temperatures, and, second, breaking of NH$_2$ bonds at higher temperatures. 

Future work is required to conclusively evaluate this hypothesis, and the possibility remains of changes in the SiN contributing to the thermal trimming results reported here. 
Additional possibilities exist of the thermal annealing of other defects in the SiN or SiO$_2$ that were not detectable in our FTIR measurements. 
Other defects can impact refractive index with an example being non-bridging oxygen hole centers and silicon dangling bonds, which are known to anneal at temperatures as low as $\approx$ 300\textdegree C \cite{Witcher2013,Neutens2018}. 

\section{Estimates of thermally induced waveguide aging}
\label{sec:5:aging}

PICs packaged in close proximity to or co-integrated with electronic integrated circuits operate at elevated temperatures defined by the power dissipation of both the electronic and photonic circuits as well as the surrounding thermal environment. Additionally, localized hot spots may form around active photonic components --- as in suspended heaters --- further increasing the required operating temperature of photonic components. As thermal annealing of materials is considered a form of accelerated thermal aging \cite{Erdogan1994,Williams1995,Kannan1997,Nemilov2000}, the observed changes in the effective index of the SiN waveguides due to thermal trimming are expected to also occur at lower temperatures --- but over a longer period of time. In this section, we estimate the aging behavior of the SiN waveguides at 125 to 250\textdegree C temperatures via extrapolations of the trimming characterization data in Section \ref{sec:3:1:trimming}. Though these estimates rely on measurements of suspended heater structures, they are independent of the source of heat and may also be relevant to SiN waveguides without suspended heaters.

\begin{figure}[!t]
    \centering
    \includegraphics[width=0.92\textwidth]{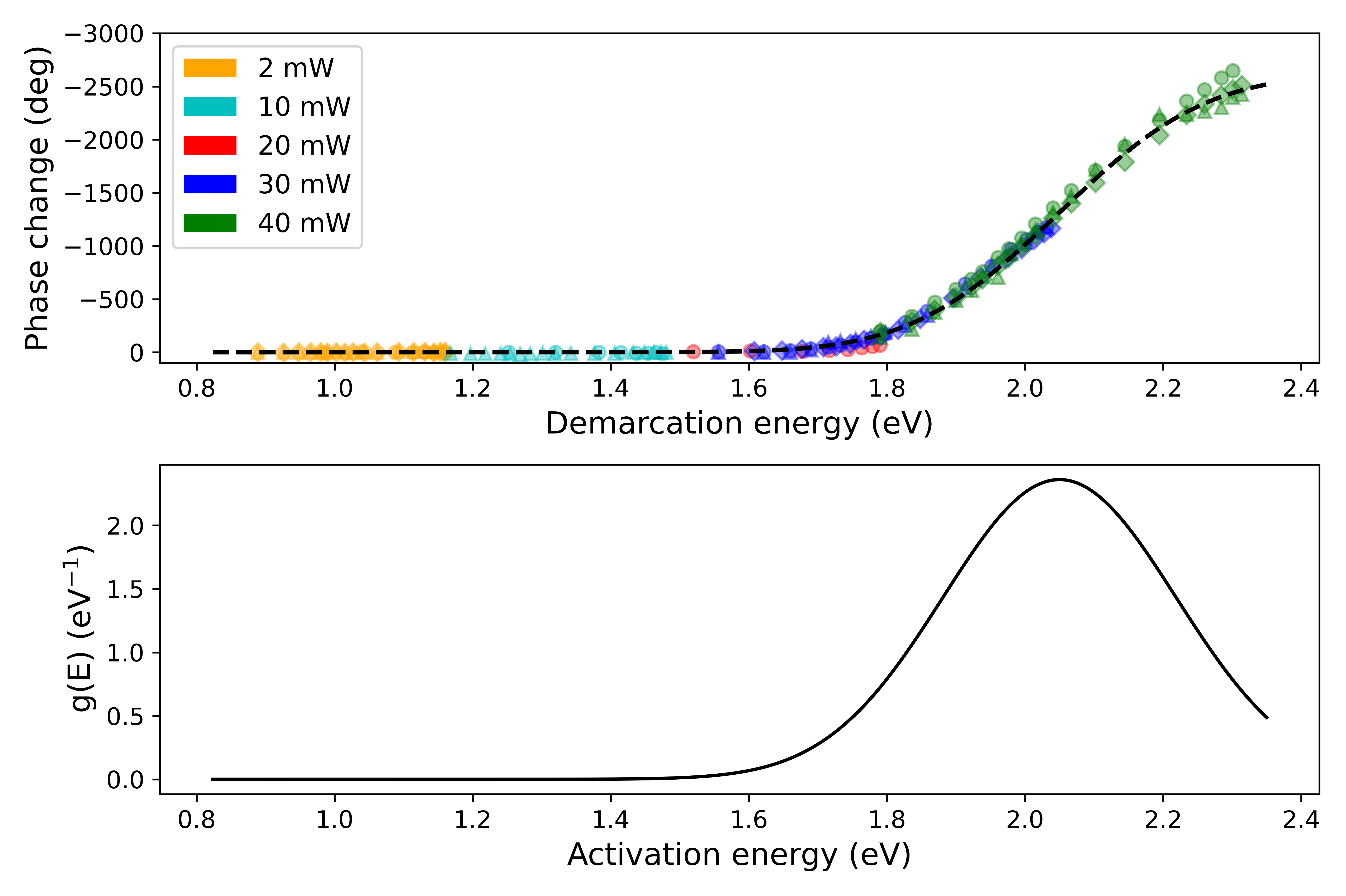}
            \caption{(a) Phase change due to thermal trimming at various applied powers vs. demarcation energy ($E_d$). Points: measured data (Section \ref{sec:3:1:trimming}) with temperatures estimated from simulations and $E_d$ defined by Eq. \ref{eq:Ed} with a fitted attempt frequency ($\nu_0$). Measured data: 3 samples (trimmed devices) at 2 mW, 2 samples at 10 mW, 1 sample at 20 mW, 3 samples at 30 mW, 3 samples at 40 mW; circles, diamonds, and triangles denote different samples at the same applied power. Black dashed line: fitted master curve with an assumed Gaussian distribution of activation energies. (b) Distribution of defects corresponding to the fitted master curve in (a).}
            \label{fig:Ed}
\end{figure}

For a population of defects distributed over a range of activation energies ($E_a$) in an amorphous material, the depletion of these defects due to thermal annealing is often modeled by the progressive increase in the demarcation energy ($E_d$), which approximates that defects with $E_a < E_d$ are annealed and all defects with $E_a > E_d$ remain \cite{Erdogan1994}. 
The demarcation energy for an anneal time of $t$ at an absolute temperature of $T$ is defined as 
\begin{equation}
\label{eq:Ed}
E_d = k_BTln(\nu_0t)
\end{equation}
where $k_B$ is the Boltzmann constant and $\nu_0$ is the attempt frequency. The measured trimmed phase was related to $E_d$, Fig. \ref{fig:Ed}(a), as follows. For each applied power, $T$ was estimated as the average of the simulated temperature in the waveguide along the heater (with the simulated hot spot in the heater spanning a length of 430 \textmu m). Equation \ref{eq:Ed} was used to transform ($t$, $T$) to $E_d$ for each measured trimming data point, and $\nu_0$ was tuned for alignment of the trimming datasets (each corresponding to a different applied power) onto a continuous phase change vs. $E_d$ curve. The datasets spanned applied powers between 2 and 40 mW, corresponding to $T$ = 40 - 350\textdegree C. Figure \ref{fig:Ed}(a) shows the master curve synthesized from the aligned trimming datasets with $\nu_0 = 3.41 \times 10^{13}$ s$^{-1}$.

By assuming the shape of the distribution of the defects, $g(E)$, over $E_a$, the master curve was then be fitted to $C\int^{E_d}_0{g(E)} dE$, where $C$ is a proportionality constant, and the fitted curve was used to extrapolate the aging behavior at lower temperatures. Following Ref. \citenum{Hughey2004}, we assumed $g(E)$ to be a Gaussian distribution, Fig. \ref{fig:Ed}(b), and obtained a fit to the master curve with a mean activation energy of 2.05 eV with a full width at half maximum of 0.4 eV. We note that the activation energy estimated from this procedure is within the range of prior reports for SiOH bonds \cite{Zhuralev2000} --- consistent with the hypothesized trimming mechanism in Section \ref{sec:4:trimming_mech}.

\begin{figure}[t!]
    \centering
    \includegraphics[width=0.62\textwidth]{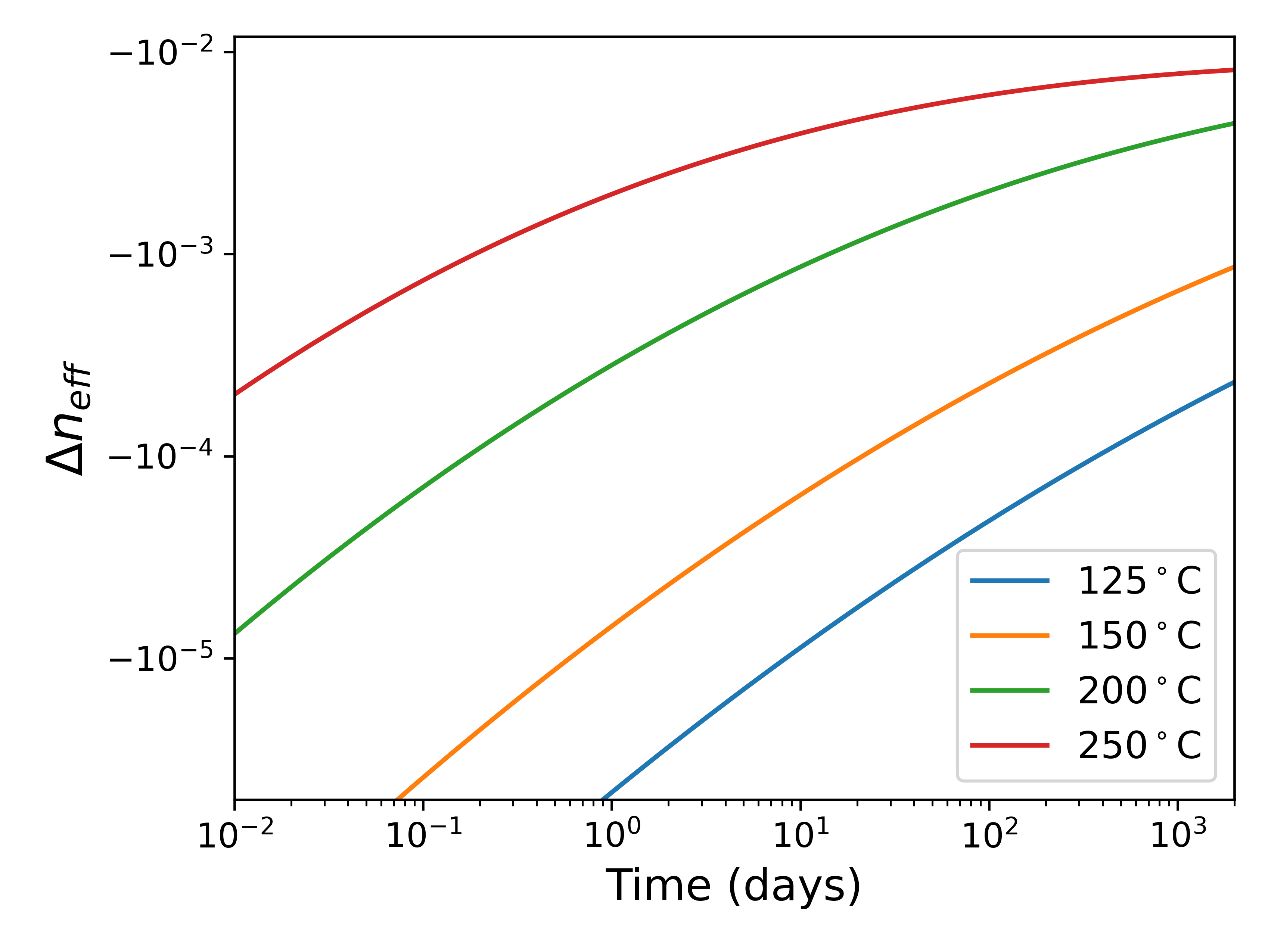}
            \caption{Estimates of the change in modal effective index of the SiN waveguides due to thermal aging at temperatures ranging from 125\textdegree C to 250\textdegree C over five years.}
            \label{fig:aging}
\end{figure}

From the fitted master curve, we estimated the aging behavior of the SiN waveguides at various operating temperatures, Fig. \ref{fig:aging}. Optical phase changes were converted to $\Delta n_{eff}$ estimates by assuming a propagation length of 430 \textmu m and $\lambda = $ 561 nm (the measurement wavelength in Section \ref{sec:3:1:trimming}). For temperatures ranging from 125 to 250 \textdegree C, $\Delta n_{eff}$ $\sim$ -1$\times 10^{-4}$ to -7.2 $\times10^{-3}$ and -2$\times 10^{-4}$ to -8.1 $\times10^{-3}$ after one and five years, respectively. The estimated $|\Delta n_{eff}|$ values due to thermal aging are comparable to those of active (e.g., thermo-optic) phase tuners and, thus, present important considerations in the design of PICs operating at elevated temperatures. For instance, photonic devices with phase tuners for stabilizing and locking bias points or channel wavelengths may require extended tuning ranges to accommodate aging. Similarly, for fixed tuning ranges, aging may set a maximum operating temperature and lifetime.

Thermal trimming offers two paths for counteracting the effects of aging. First, owing to the compatibility of our \emph{in situ} trimming method with packaged PICs, phase correcting trimming operations may be performed periodically throughout the lifetime of the PIC. However, as the sign of $\Delta n_{eff}$ is the same for both trimming and aging, such phase corrections require a 2$\pi$ rad. offset. Second, PICs may be hardened against thermal aging through annealing of sensitive structures, (e.g. folded waveguides, ring resonators) embedded within suspended heaters. This is similar in concept to previous proposals to harden UV written fiber Bragg gratings by thermal annealing \cite{Erdogan1994,Kannan1997}. Noting the saturation of $\Delta n_{eff}$ with trimming/aging time [Figs. \ref{fig:fig2}(d) and \ref{fig:aging}] --- a result of the depletion of available defects (e.g., SiOH bonds) for annealing --- extended trimming operations may be performed on photonic devices, reducing the magnitude of $\Delta n_{eff}$ possible with subsequent heat exposure. Additionally, these trimming operations may be stopped at favorable operating points for devices (compensating for variation in device dimensions). Beyond thermally induced aging, visible spectrum PICs may also be susceptible to aging from guiding highly-confined visible light. It is known that visible light can photobleach (optically anneal) defects in optical fibers \cite{Friebele:81}, fiber Bragg gratings \cite{Askins:1997}, SiN thin films \cite{Chen1994}, and SiN waveguides \cite{Neutens2018}. Extended exposure to visible (particularly blue) light may result in waveguide aging, however, future investigations are required to quantify this aging and determine whether thermal trimming may be used to harden devices against it. 

The estimates of waveguide aging given in this section are limited by the accuracy of the simulated temperatures of the suspended structure and are affected by the choice of $g(E)$ for the fit of the master curve. Design improvements to the suspended heaters for increased temperature uniformity and, also, direct temperature measurements during heater operation are expected to improve the accuracy of these aging estimates. In addition, the thermal aging behavior may change significantly if passivation layers are introduced \cite{Haque1997_passiv}. Nonetheless, the results of this section highlight the utility of suspended heaters for the testing of accelerated thermal aging of waveguides, with future work being required to determine the generality of the resulting aging estimates toward non-suspended waveguides operating at elevated temperatures. 

\section{Conclusion}
\label{sec:6:conclusion}

In summary, we have proposed and experimentally demonstrated an \emph{in situ} post-fabrication refractive index trimming method for visible-light SiN waveguides. Suspended heaters enabled high estimated local temperatures near 350\textdegree C with only 40 mW of applied electrical power, leading to effective index changes of -3.6$\times$10$^{-3}$ and -8.3$\times$10$^{-3}$ after 10 and 2738 minutes of trimming, respectively ($\lambda = $ 561 nm). The trimming approach was applied to thermo-optic MZI switches with a multi-pass phase shifter design to compensate for fabrication variations, reducing the bias power from $0.29-2.32$ mW in as-fabricated devices to $0.1-0.16$ mW after trimming. Suspended heaters with SiN waveguides enable both low-power phase tuning and medium-power thermal trimming --- a convenient combination for reducing power consumption in photonic integrated circuits and relaxing drive circuitry requirements for \emph{in situ} post-fabrication thermal trimming. The devices were fabricated in our visible-light integrated photonics platform on 200-mm Si wafers, adding to the growing list of optical functionalities available for visible-light photonic integrated circuits. Through characterization of the SiN waveguide core and SiO$_2$ cladding materials with and without thermal treatment, we have identified refractive index changes in the SiO$_2$ cladding as a possible mechanism of the thermal trimming reported here --- providing a foundation for future investigations and engineering of this trimming method. Finally, with extrapolations of our measured trimming data and the demarcation energy approximation, we estimated the thermal aging behavior of the SiN waveguides at 125 to 250\textdegree C operating temperatures. This exploration highlights the importance of aging considerations in photonic circuits operating at $\gtrsim$ 100\textdegree C temperatures over extended periods, while also identifying suspended heaters as a possible means of hardening devices against aging and studying these effects via accelerated aging. 

\clearpage

\section*{Appendix}

\begin{figure}[h]
    \centering
    \includegraphics[width=0.73\textwidth]{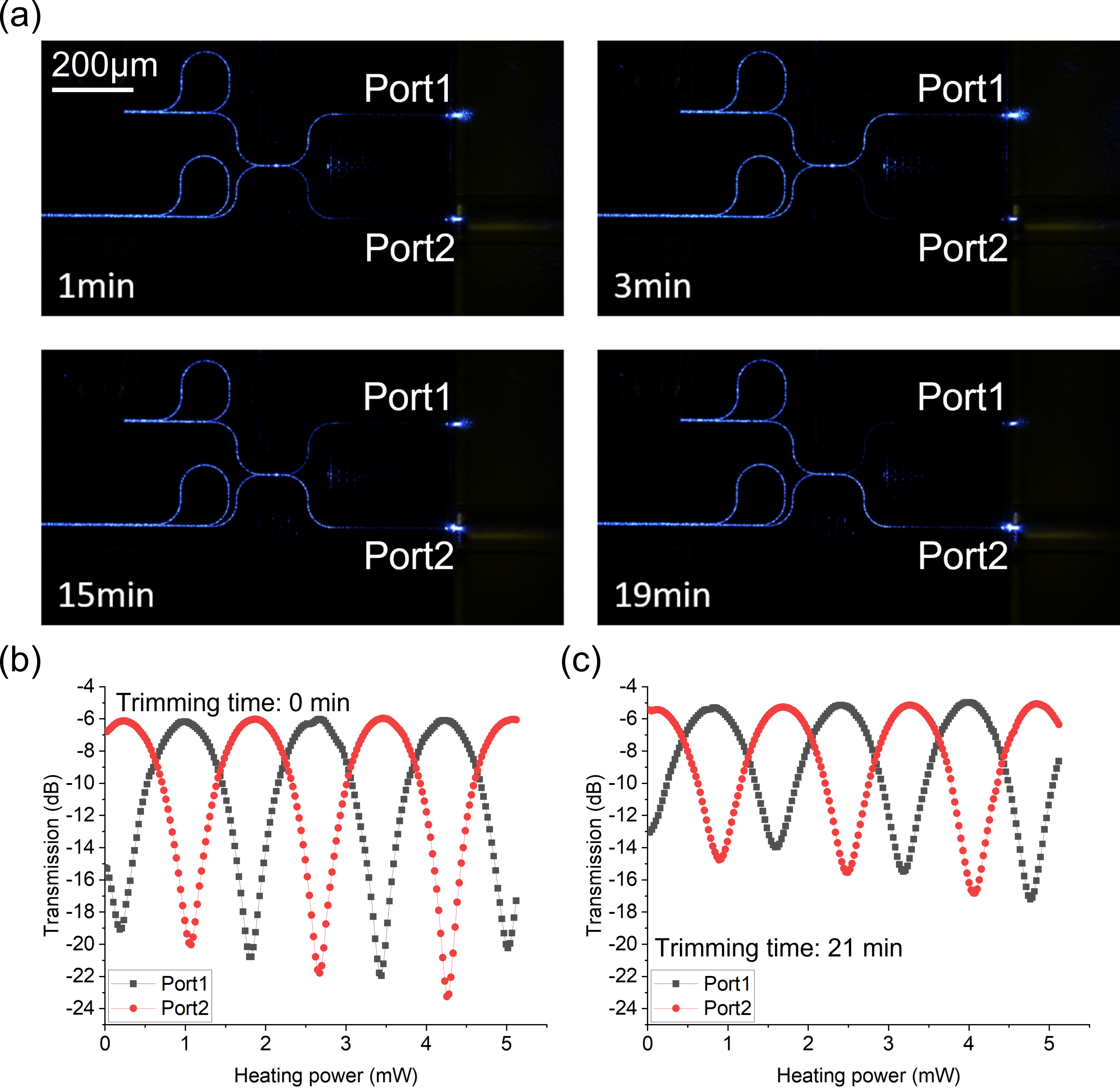}
            \caption{Thermal trimming of a thermo-optic MZI switch at $\lambda = 445$ nm. (a) Optical micrographs of the outputs of the MZI following trimming for 1, 3, 15, and 19 minutes. (b), (c) Transmission versus heating power (b) before trimming and (c) after 21 minutes of trimming.}
            \label{fig:fig6}
\end{figure}

To investigate the trimming characteristics at shorter wavelengths, we carried out measurements at $\lambda = $ 445 nm using a single-wavelength laser (Coherent OBIS LX). As in Section \ref{sec:3:2:trim2bias}, a 1$\times$2 MZI thermo-optic switch was thermally trimmed to adjust its bias point. The MZI was similar to Fig. \ref{fig:fig3}(a) but with different MMI coupler designs for operation at $\lambda = $ 445 nm. Optical micrographs of the MZI at various times during the trimming process are shown in Fig. \ref{fig:fig6}(a). 36 mW of power was applied to the TiN heater. From 1 min to 19 min of trimming, minimal/maximal output power in Port2 was obtained at 3 and 19 min, respectively. The transmission versus heating power traces before and after trimming are shown in Figs. \ref{fig:fig6}(b) and \ref{fig:fig6}(c), respectively. The device was slightly over-trimmed with 21 min of total trimming time, missing the trimming time required to reduce the bias power. Nonetheless, the data in Fig. \ref{fig:fig6} demonstrate that the thermal trimming method is also compatible with blue light.

\begin{backmatter}
\bmsection{Funding}
Max-Planck-Gesellschaft
\bmsection{Acknowledgments}
This work was supported by the Max Planck Society. The authors thank Dr. Wenqiong Tang of EAG laboratories for conducting the FTIR measurements, and Dr. Chandrasekhar Naisa at the Max Planck Institute of Microstructure Physics for assistance with operating the tube furnace. The authors also thank Christian Patzig and Thomas Höche at the Fraunhofer-Institut für Mikrostruktur von Werkstoffen und Systemen IMWS for transmission electron microscopy and energy dispersive X-ray measurements.
\bmsection{Disclosures}
The authors declare no conflicts of interest.
\bmsection{Data availability}
Data underlying the results presented in this paper are not publicly available at this time but may be obtained from the authors upon reasonable request.
\end{backmatter}

\end{document}